\documentclass[conference]{IEEEtran}
\usepackage{subfigure,graphicx,comment,cite}
\usepackage{amsmath,amssymb}
\usepackage{algorithmic}
\usepackage{psfrag}

\newcommand{\inst}[1]{$^{#1}$}

\newcommand{\N}{N}
\newcommand{\PA}{\ensuremath{\mathcal{P}}}
\newcommand{\SE}{\ensuremath{\mathcal{S}}}
\newcommand{\source}{\ensuremath{s}}

\newcommand{\abil}{Abilene}
\newcommand{\geant}{Geant}
\newcommand{\tiger}{Tiger2}

\newcommand{\path}[1]{\ensuremath{\mathcal{#1}}} 
\newcommand{\lenP}{\ensuremath{L_p}} 
\newcommand{\lenS}{\ensuremath{L_s}} 
\newcommand{\degr}{\ensuremath{\bar{\delta}}} 

\newcommand{\nadvertise}{\ensuremath{M_a}} 
\newcommand{\nrefresh}{\ensuremath{M_r}} 
\newcommand{\nm}{\nadvertise} 
\newcommand{\np}{\nrefresh}  

\newcommand{\pc}{\ensuremath{C_p}}   
\newcommand{\seco}{\ensuremath{C_s}} 
\newcommand{\po}{\ensuremath{O_p}}   
\newcommand{\so}{\ensuremath{O_s}}   

\newcommand{\Be}{\ensuremath{\beta}} 
\newcommand{\ML}{\ensuremath{\kappa}} 
\newcommand{\PB}{\ensuremath{p_b}} 

\newcommand{\tadvertise}{\ensuremath{\tau_{a}}} 
\newcommand{\trefresh}{\ensuremath{\tau_{r}}} 
\newcommand{\tm}{\tadvertise} 
\newcommand{\tr}{\trefresh} 

\newcommand{\FW}{\text{ADV}}

\newcommand{\nid}{\ensuremath{ID}}

\title{Adaptive Probabilistic Flooding\\ for Multipath Routing}
  \author{%
   \IEEEauthorblockN{Christophe Betoule\inst{2}, Thomas Bonald\inst{1}, Remi Clavier\inst{2}, Dario Rossi\inst{1}, Giuseppe Rossini\inst{1,\dagger}, Gilles Thouenon\inst{2}}\\
   \IEEEauthorblockA{\inst{1} Telecom ParisTech, Paris, France\\
   firstname.lastname@telecom-paristech.fr ($\dagger$ corresponding author)\\}
   \IEEEauthorblockA{\inst{2} Orange Labs, Lannion, France\\
   firstname.lastname@orange-ftgroup.com\\}
  }

\begin{document}

\maketitle

\begin{abstract}
In this work, we develop a distributed source routing algorithm for topology discovery suitable for ISP transport networks, that is however inspired by opportunistic algorithms used in ad hoc wireless networks. We propose a plug-and-play control plane, able to find multiple paths toward the same destination, and introduce a novel algorithm, called adaptive probabilistic flooding, to achieve this goal. By keeping a small amount of state in routers taking part in the discovery process, our technique significantly limits the amount of control messages exchanged with flooding -- and, at the same time, it only minimally affects the quality of the discovered multiple path with respect to the optimal solution. Simple analytical bounds, confirmed by results gathered with extensive simulation on four realistic topologies, show our approach to be of high practical interest.   
\end{abstract}

\section{Introduction}\label{sec:Introduction}

Reducing the overall power consumption of the Internet is a major challenge for future networking technologies. It is commonly accepted that a significant amount of power is consumed by the IP address look-up algorithm, which becomes prohibitive with the growing size of the routing tables. A viable option for telecom carriers and ISPs is to replace IGP routing by some appropriate level-2 technology, thus forming 
full mesh subnetworks at the IP level.
While Ethernet is a natural candidate, it still requires highly dynamic switching tables; moreover, path discovery is based on flooding, which doesn't make it scalable.

In this paper, we focus on a novel level-2 architecture 
that relies on the following key principles:
\begin{itemize}
\item each source maintains a set of multiple distinct paths to each destination;
\item the header of each packet consists of a sequence of labels, one per node on the corresponding selected path;
\item each node receiving  a packet  (i) removes the first label of the header, if any, and forwards the packet accordingly (the node is a relay) (ii) sends  it to the IP layer in the absence of label (the node is the destination).
\end{itemize} 
Note that core nodes do {\it not} maintain routing tables, which avoid expensive address lookup algorithms. All routing information is contained in the packet itself. Paths are built and updated by the source nodes, which enables {\it multipath} routing. For instance, each source maintains a primary path to each destination (typically, the shortest path), as well as a secondary path in case of failure or traffic surge. Unlike traditional IP routing, this secondary path  is always available, which enables fast restoration in case of failure. 

A critical component of this architecture is the algorithm used to discover paths. We propose a novel 
  flooding algorithm inspired by the opportunistic algorithms of peer-to-peer applications, which consists in adapting the flooding rate of core nodes to the number of already received discovery messages. 
We refer to this algorithm as {\it adaptive probabilistic flooding}. At the cost of a limited number of state variables  in core routers, the algorithm is able to discover multiple paths in a quasi-optimal way. Using analytical bounds, we prove that, unlike pure flooding, the algorithm scales with the network size.

The rest of the paper is organized as follows. Related work is presented in  the next section. 
The path discovery algorithm is  described in Section   \ref{sec:Definitions}.
Section \ref{sec:Perf} is devoted to the performance analysis. Section \ref{sec:Future} concludes the paper.

\section{Related work}\label{sec:Related}

Routing is a critical component of the Internet, and as such has long been studied by the scientific community. The problem of finding a \emph{single path} interconnecting any two nodes of a graph is solved by well-known  algorithms like Dijkstra and Bellman-Ford, which have been implemented in widely deployed protocols such as OSPF and RIP, respectively. However, interconnecting nodes through a   single path (typically, the shortest) does not make the network resilient against  failures and traffic surges. Hence, different techniques relying on \emph{multiple paths} have been proposed. For instance, ECMP~\cite{HoppsRfc2992} aims at balancing load  over multiple paths of equal cost.   In standard IP/MPLS networks,  the control and data planes are generally considered jointly; multipath routing  is then achieved through a centralized algorithm, solving some  standard  multicommodity flow problem~\cite{Guerin97Globecom,Srihari01Iwqos}. 

In this work, we focus on the control plane and address the issue of the efficient discovery of  multiple paths, as in~\cite{Ogier93IEETrans,Eppstein94Fcs, Sidhu91Sigcomm,Merindol09Infocom,Johnson07Rfc4728,Blesa04Evo,Chen02ANW,Chen98Infocom}. Some of the above papers consider the problem of finding a pair of disjoint paths between any two nodes within the network, considering that the primary (shortest) path is \emph{already known}. In~\cite{Ogier93IEETrans,Eppstein94Fcs, Sidhu91Sigcomm}, the problem is shown to boil down  to a standard shortest path problem by some appropriate modification of the original graph.
Ogier, Rutenburg and Shacham define  in \cite{Ogier93IEETrans} an algorithm that achieves such a graph modification and   evaluates its performance in terms of communication, convergence time, and space complexity. 
Sidhu, Nair and Abdallah enhance this algorithm in  \cite{Sidhu91Sigcomm}  by finding all possible disjoint paths between any two nodes of the network. 
Eppstein proposes in \cite{Eppstein94Fcs} an algorithm that finds the first $k$ shortest paths between any two nodes by a breadth first search of a 4-heap in which every node represents a path. 
Finally, works like~\cite{Merindol09Infocom} take a more practical approach, and enhance IP routing by means of centralized algorithms to determine disjoint paths, and distributing such paths through the routers taking care of avoiding loops. 

The problem of finding multiple paths {\it without} any a priori knowledge such as the shortest path is quite different.
Most algorithms then rely on flooding  \cite{Johnson07Rfc4728,Blesa04Evo,Chen02ANW,Chen98Infocom}.
A swarm intelligence based solution is proposed with  the Ant Colony Optimization (ACO) algorithm~\cite{Blesa04Evo}, where a set of ants is spread through the network in order to discover disjoint multiple paths: the pheromone left by the ants is employed in order to avoid already crossed paths.  In \cite{Chen02ANW}, a flooding algorithm on layered routing architecture is employed: the basic idea is to give a score to each packet and to decrease this score for each link on which the packet is flooded; only nodes along the best path can increase the score and re-flood the packet. Authors in~\cite{Chen98Infocom} propose a strategy where a scout message walks through the network accumulating nodes discovered in the walk: nodes re-flood the message only when  the current discovered path differs significantly from the stored  shortest path, allowing thereby to find multiple paths. 
In a very different context, namely ad-hoc wireless networks, flooding-based technique are exploited by  Dynamic Source Routing (DSR) \cite{Johnson07Rfc4728} where however multiple paths are not taken into account.  

To the best of our knowledge, our approach introduces a number of novel ingredients. Similarly to DSR~\cite{Johnson07Rfc4728}, in our approach the source inserts one label per node on the path to the destination, and the data plane forwards the packet by popping a label from the packet header at each hop.  Unlike DSR, we don't have collisions, or problems inherent to wireless networks, yielding to a radically different algorithm design.
The adaptive  probabilistic algorithm we propose greatly limits the number of exchanged messages, as in \cite{Chen02ANW}, without however requiring packets to carry the scores associated with the discovery algorithm.
The limited amount of state kept by nodes is not used  to avoid crossing already traveled paths as in~\cite{Blesa04Evo}, but  rather to avoid taking these paths too often, which has important consequences on the quality of the multiple paths discovered.
Finally, unlike~\cite{Chen98Infocom}, the algorithm does not rely on  threshold-based decisions, nor it depends on topological properties of the network -- rather, its design makes it robust and auto-terminating irrespectively of its actual parameter setting, with performance that degrades gracefully in case of parameter misguidance.

\section{Path discovery algorithm}\label{sec:Definitions}

\subsection{Overview}

We aim at designing  a distributed algorithm for path discovery, capable of finding multiple, possibly disjoint paths between any pairs of nodes. To do so, each node  periodically advertises its presence  by means of some flooding procedure described below. Specifically, each node sends  an \emph{advertisement} message  every \tadvertise\ seconds; typical values range from a few seconds to minutes \cite{Srihari01Iwqos}. Besides,  each node sends keep-alive messages over each path in order to ensure that this path has not failed; the corresponding \emph{refresh} messages are sent every  \trefresh\ secondes, typically set to a few milliseconds \cite{Fouli09Ieeecm}.

During the flooding procedure, each relay node adds its identifier to the advertisement messages it receives, so that these messages carry information concerning the whole traveled  path. Upon reception of an advertisement message, a node learns a path from the source of this message, as well as from any intermediate node on this path, as in DSR~\cite{Johnson07Rfc4728}.
Flooding decisions are taken independently by each node, and consitute the core of the algorithm. The main idea is that nodes need to flood a received message \emph{at least once}, so that shortest paths are discovered. Nodes actually need to flood the message \emph{multiple times}, in order to discover further paths beyond the shortest one. The number of flooding decisions   is critical with respect to both the quality of the path discovery and the overhead of the algorithm.

A simple option could consist in including a Time To Leave (TTL) field in the packet, so as to   interrupt the flooding process when some pre-configured maximum path length is reached.  The selection of a proper TTL value is  critical in this case: if the TTL is shorter than the graph diameter $D$ for instance, then connectivity cannot be guaranteed; if the TTL is too large, the overhead of the algorithm becomes prohibitive (as the number of relayed messages is exponential in the TTL).

We propose an alternative approach based on {\it adaptive probabilistic flooding}. Any node receiving some advertisement message from source $s$ floods this message the first time, and floods it with some decreasing probability the following times. Specifically,  node $i$ floods an advertisement message generated by source node $s$ over all its links  (except the one from which  it has received the message) with probability:
\begin{equation}
  P=\beta^{n_{i,s}}
  \label{eq:p}
\end{equation}
where $\beta$ is some fixed parameter and $n_{i,s}$ is a counter, stored at node $i$, of the number of times node $i$ has already received an advertisement originated by node $s$. The flooding decision are taken independently on each link, and the counter is reset periodically, as explained later. 
Note that node  $i$ floods the first advertisement message it receives for source node $s$ since $n_{i,s} = 0$  in this case. As further messages are received, flooding will become exponentially less likely, according to the backoff parameter $\beta$. 
The quality of the path discovery is expected to increase with $\beta$, at the expense of larger overhead. 
However, we shall see that performance is not very sensitive to this parameter, which makes the algorithm robust and pratically interesting.

\subsection{Primary and secondary paths}
Consider a network, modeled as an undirected graph $G=(E,V)$, composed of $|V|=\N$ routers, in which any pair of adjacent routers are connected by a single link for simplicity (the algorithm can be easily extended to  the general case of multiple links between any pair of nodes). Between any two routers $i,j\in V$, we are interested in finding a pair of \emph{paths}, i.e., sequences of edges 
connecting node $i$ to $j$.  We denote by  \PA\ and \SE\ the primary and secondary paths, respectively, returned by the adaptive probabilistic algorithm on graph $G$, as described below. 
We denote by \lenP\ and \lenS\ the respective lengths of these paths.

To gauge the quality of the primary and secondary paths found by our algorithm, 
we need to define target path properties. The primary path is expected to be  the shortest path in number of hops; in other words, we say that  \PA\ is optimal if it belongs to the set of shortest paths from $i$ to $j$  in $G$ (as there may be several such paths). The secondary path is expected to  minimize the similarity with the primary path, $\PA \cap \SE$. Note that this choice reduces the share of faith between these paths, improving network resilience against failures and traffic surges.

To find the optimal secondary path \SE, we consider a modified graph $G'$ in which the cost of links along the primary path \PA\ are increased by the network diameter~\cite{Ogier93IEETrans}, and other link costs are unitary. As links belonging to \PA\ are now discarded due to higher cost, running Dijkstra on $G'$ we retrieve a path \path{S'} minimizing the similarity function $\PA\cap \path{S'}$ (notice that since nodes along the primary path are not removed from $G'$, they can be included in \path{S'} only if strictly necessary as the path would otherwise be disconnected). We say that the secondary path found by the algorithm \SE\ is  optimal if $|\path{P}\cap \path{S}|=|\path{P}\cap \path{S'}|$ and $\lenS=L_{s'}$, i.e., the length \lenS\ of the secondary path is equal to the length $L_{s'}$ of the optimal  $\path{S'}$ (as there may be multiple disjoint paths minimizing the similarity with the shortest path).

\subsection{Metrics}

To precisely quantify the  \emph{overhead} vs \emph{path quality} tradeoff early outlined, we resort to the following metrics.
The average amount of messages handled by any given node during the advertisement procedure is denoted with \nm. For each path, the refresh process then requires each node to send keep-alive messages: we denote by  \np\ the average number of such messages (counted once per every link traveled). The relative weigth of $\nm /(\np +\nm)$ quantifies the \emph{overhead} induced by the advertisement process.

In the following, we evaluate the algorithm in terms of \emph{connectivity} along the primary and secondary path (i.e., whether paths  \PA\ and \SE\ joining any two nodes $i,j\in V$ exist) and \emph{optimality} (i.e., whether  \PA\ and \SE\ are optimal according to the above definitions).
We  express connectivity in terms of the probability \pc\ (respectively, \seco) that, $\forall i,j\in V$, nodes $i$ and $j$ are connected by some primary (respectively, secondary) path.   
We express  optimality in terms of the probability \po\ (respectively, \so) that the primary path is also the shortest (respectively, that the secondary path is the shortest, most diverse path).


\subsection{Pseudocode}

\begin{figure}[t]
\begin{center}\small
\begin{tabular}{|l|}\hline
\begin{minipage}{0.9\hsize}
\vspace{3mm}
{\small
\begin{algorithmic}[1]
\WHILE{ \{receiving message \FW\} }
	\STATE $\ell \leftarrow length(\FW.\nid)$ 
	\FORALL{ \{ $i \in [0,\ell]$ \} }    
		\IF{ \{$\FW.\nid[i] = j$\} }			
			\STATE exit
			\COMMENT{ Break loop and abort flooding }
		\ELSE
			\STATE d $\leftarrow$ \FW.\nid[i] \COMMENT{ Destination }
			\STATE $\mathcal{L}_{j,d}$$\leftarrow$$(\FW.\nid[\ell], \ldots , \FW.\nid[$i$])$\\
			\COMMENT{ Overhearing advertised paths from \FW}; 	

			\IF{  \{ $\nexists \path{P}_{j,d} \vee length(\path{L}_{j,d})<length(\path{P}_{j,d})$\} }				\STATE $\path{P}_{j,d} \leftarrow \path{L}_{j,d}$ 
				\COMMENT{ Update primary path}
			\ENDIF
			\IF{ \{$ \nexists \path{S}_{j,d} \vee |\path{P}_{j,d}\cap \path{L}_{j,d}| < |\path{P}_{j,d} \cap \path{S}_{j,d}|   \vee (|\path{P}_{j,d}\cap \path{L}_{j,d}| = |\path{P}_{j,d} \cap \path{S}_{j,d}|  \wedge length(\path{L}_{j,d})< length(\path{S}_{j,d}))$\} }
				\STATE $\path{S}_{j,d} \leftarrow \path{L}_{j,d}$ 
				\COMMENT{ Update secondary path}
			\ENDIF
			
		\ENDIF
	\ENDFOR
	\STATE {append $j$ to $\FW.\nid$ }
	\STATE s $\leftarrow$ \FW.\nid[0]  \COMMENT{ Source }	
			
	\FORALL {\{$next\in  neighbors(j)$ \}}
		\IF{ \{ $next \neq \FW.\nid[\ell-1]$   \} }						\STATE {send \FW\ to $next$ w.p. $\beta^{n_s}$} 	\\
			\COMMENT{ Adaptive probabilistic flooding} 
		\ENDIF
	\ENDFOR
	\STATE n$_s$++
	\COMMENT{ Update counter associated with source $s$} 
\ENDWHILE
\end{algorithmic}
\vspace{1mm}
}

\end{minipage} \\ \hline
\end{tabular}
\end{center}
\caption{Algorithm pseudo-code for a generic node $j$ of the network}
\label{fig:alg}
\end{figure}

A pseudocode description of the algorithm is  given in Fig.~\ref{fig:alg}.
A source node $s$ initiates the advertisement process by flooding an advertisement packet \FW\ to all its neigbors. 
The flooded packet contains a  list of node identifiers \nid, initially set to \nid[0]=\source\ by the source, to which each node appends its own identifier. 
Upon reception of an advertisement  packet \FW, a node learns a (backward) path to the source $s$ and to any  intermediate node $d = \FW.\nid[i]$ along the path.  In case the receiver $j$ detects a loop (finding its identifier within the \nid\ list), it discards the message and aborts the flooding procedure. Otherwise, it analyzes, and possibly stores, the newly learned path $\path{O}_{j,d}$. Specifically, the primary (and secondary) path is first set if not existent yet. Also, if the newly overheard path is shorter than the primary path $length(\path{L}_{j,d})<length(\path{P}_{j,d})$, then  the primary path is updated with the overheard one. Similarly,  if the overheard path has lower similarity than the current secondary path $|\path{P}_{j,d}\cap \path{L}_{j,d}| < |\path{P}_{j,d} \cap \path{S}_{j,d}|$, or if it has equal similarity but is shorter than the secondary $|\path{P}_{j,d}\cap \path{L}_{j,d}| = |\path{P}_{j,d} \cap \path{S}_{j,d}|  \wedge length(\path{L}_{j,d})< length(\path{S}_{j,d})$, then the secondary path is updated. 

Notice that we expect messages on the shortest path to reach a node \emph{before} messages that take longer paths: in case of homogeneous setup (i.e., equal latencies) this invariant always holds. As such, the shortest path is chosen relatively early in the advertisement process and is typically never changed later on, as its change would also affect the quality of the secondary path. Not shown in the pseudocode for the sake of clarity,  ties in the secondary path selection are broken at random. 

Finally, after having added its own identifier to the \FW.\nid, the node probabilistically  floods the \FW\ message, with independent decisions per each neighbor (except the node \FW.\nid[$\ell$-1] from which the message came), and update the per-source counter $n_s$. As we shall see, the duration of each discovery phase is much shorter than the discovery period \tadvertise, which implies that the counter $n_s$ can be safely reset for new discoveries.

\section{Performance evaluation}
\label{sec:Perf}

\begin{figure}[t]
\begin{center}
\includegraphics[width=0.35\textwidth]{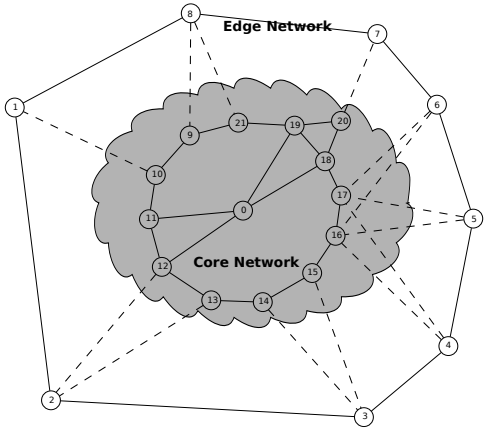}
\end{center}
\caption{\tiger: an example of hierarchical AS network.  The core network is highlighted by the grey balloon, and router 0 represents the collect point toward the Internet core.}
\label{fig:tiger}
\end{figure}

In this section, we first evaluate the overhead of the algorithm through a simple analytical model, and then employ discrete event simulation to validate the  analysis and evaluate performance in 
 terms of the quality of discovered paths.

Simulations are carried on with Omnet++~\cite{Varga10Omnet} over four
different network topologies, whose  most significant properties are summarized
in Tab.~\ref{tab:prop}.  
Specifically, Tab.~\ref{tab:prop} reports the number of nodes $N$, the
average and standard deviation of the node degree, $\degr$ and $\sigma$,	the average length of the optimal primary $\overline{\lenP}$ and secondary $\overline{\lenS}$ paths, the diameter $D$ of the original grap $G$ and the largest diameter $D'$ over the modified $G'$ graphs. 

 Note that we consider both real network topologies (Tiger2~\cite{tiger2}, Abilene~\cite{abilene}, Geant~\cite{geant}), corresponding to different segments,  as well as a set of 50 synthetic random graphs (Random) whose number of nodes and degree distribution loosely fit the real topologies (in case of the Random topology,  $D$ and $D'$ are averaged over the 50 considered instances).
All network topologies are well known except for Tiger2, that we depict in Fig.~\ref{fig:tiger}. For the time being, we use homogeneous settings (i.e., constant and equal delay on every link), and no failures happen within the network. The evaluation of more complex (heterogeneous delay, failures, etc.) scenarios is part of our ongoing work.

\begin{table}[t]
\begin{center}
\caption{Topological properties of the network scenarios}
\label{tab:prop}
\begin{tabular}{llccccccc}
Network &  Segment   &  N  &   \degr  &     $\sigma$  &     $\overline{\lenP}$  &     $\overline{\lenS}$  &     $D$  &     $D'$  \\
\hline
\tiger~\cite{tiger2}  &metro       &  22  &  3.6  &  0.6  &  2.7  &  3.7  &  5     &  6     \\
\geant~\cite{geant}  &aggregation &  22  &  3.4  &  1.4  &  2.6  &  4.0  &  6     &  10    \\
\abil~\cite{abilene}   &core        &  11  &  2.6  &  0.5  &  2.4  &  4.2  &  5     &  8     \\
Random  &synthetic   &  22  &  3.2  &  1.6  &  2.6  &  3.8  &  5.4   &  7.3   \\
\hline
\end{tabular}
\end{center}
\end{table}	


\begin{figure*}[t]
\begin{center}
\subfigure[\label{fig:msg}Number of advertisement messages generated by a full discovery round of all the nodes in the network as a function of \Be]{\includegraphics[width=0.3\textwidth]{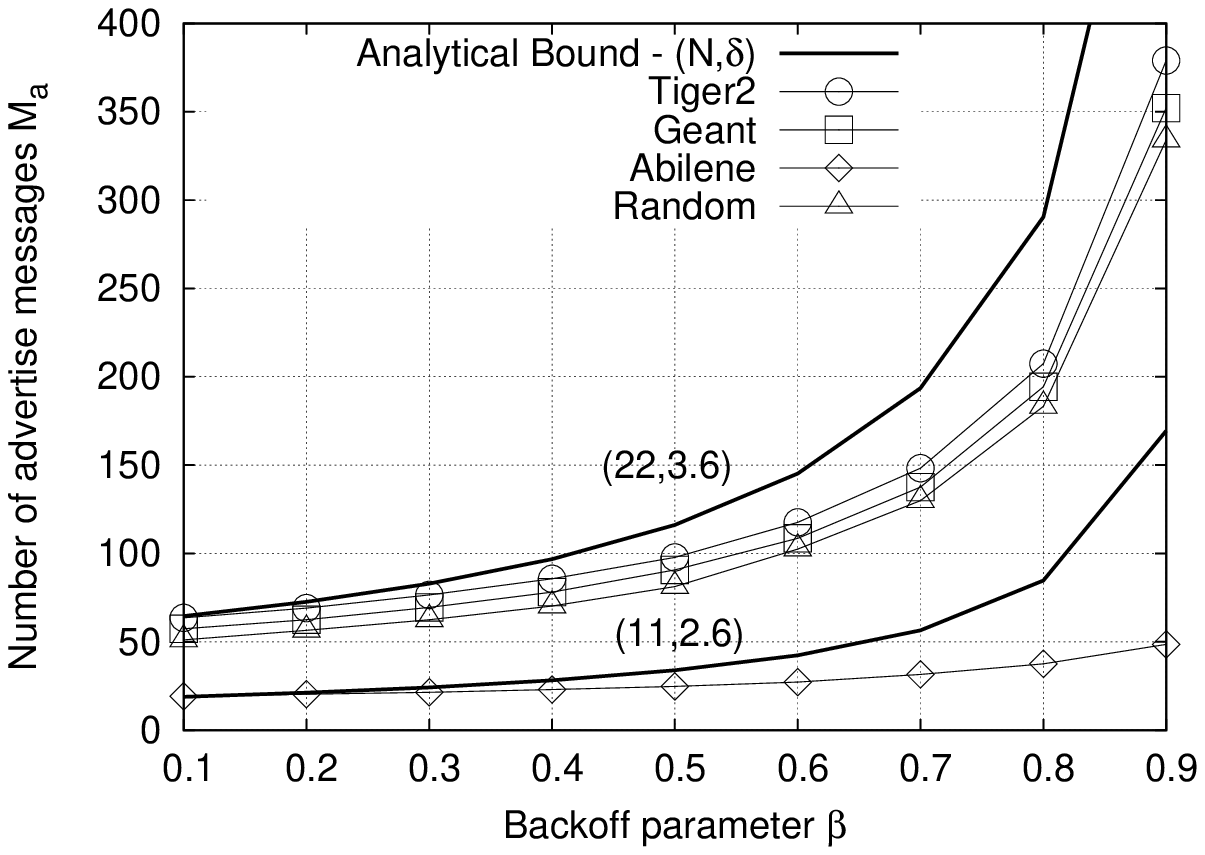}}
\subfigure[\label{fig:overhead}Number of refresh \nrefresh\ and avertisemed \nadvertise\ messages generated during a fixed time window $W=1$\,sec as a function of the advertisement period \tadvertise\ (Tiger2 network)]{\includegraphics[width=0.3\textwidth]{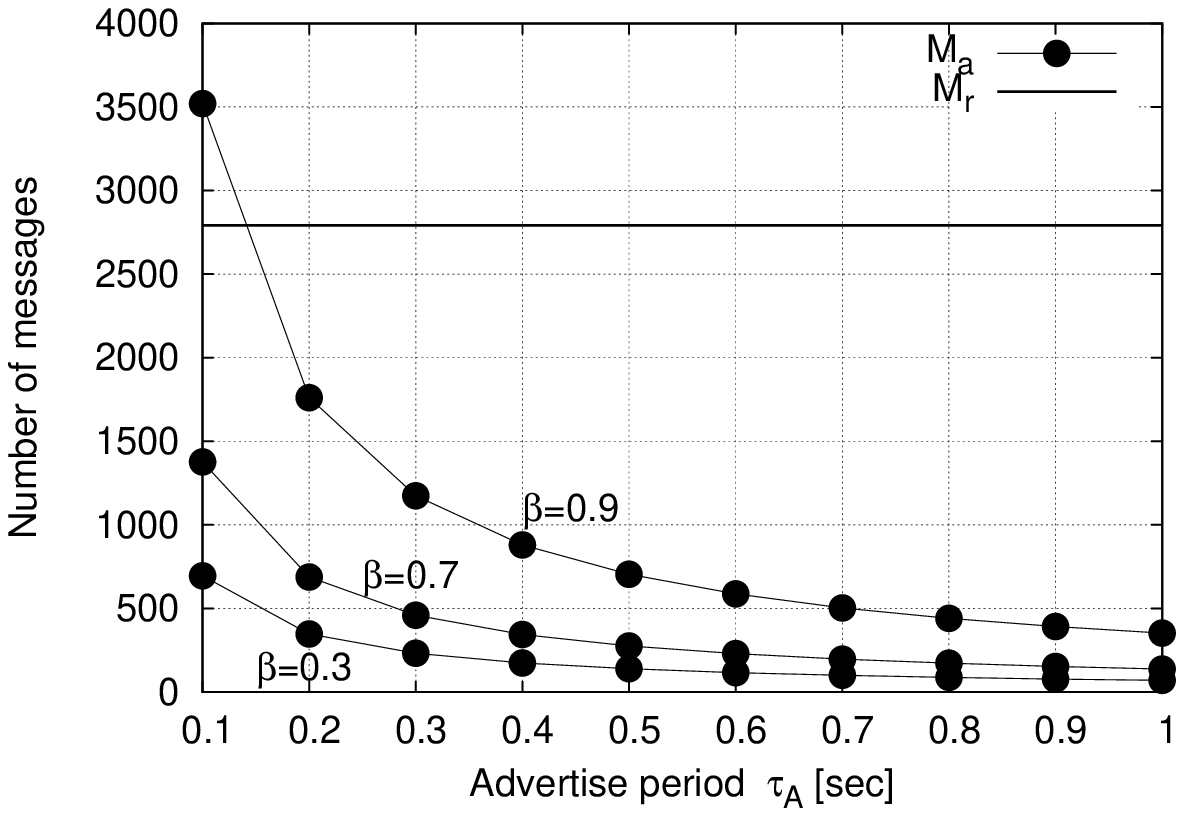}}
\subfigure[\label{fig:msgt}Time evolution of the number of avertisement messages (Tiger2 network)]{\includegraphics[width=0.3\textwidth]{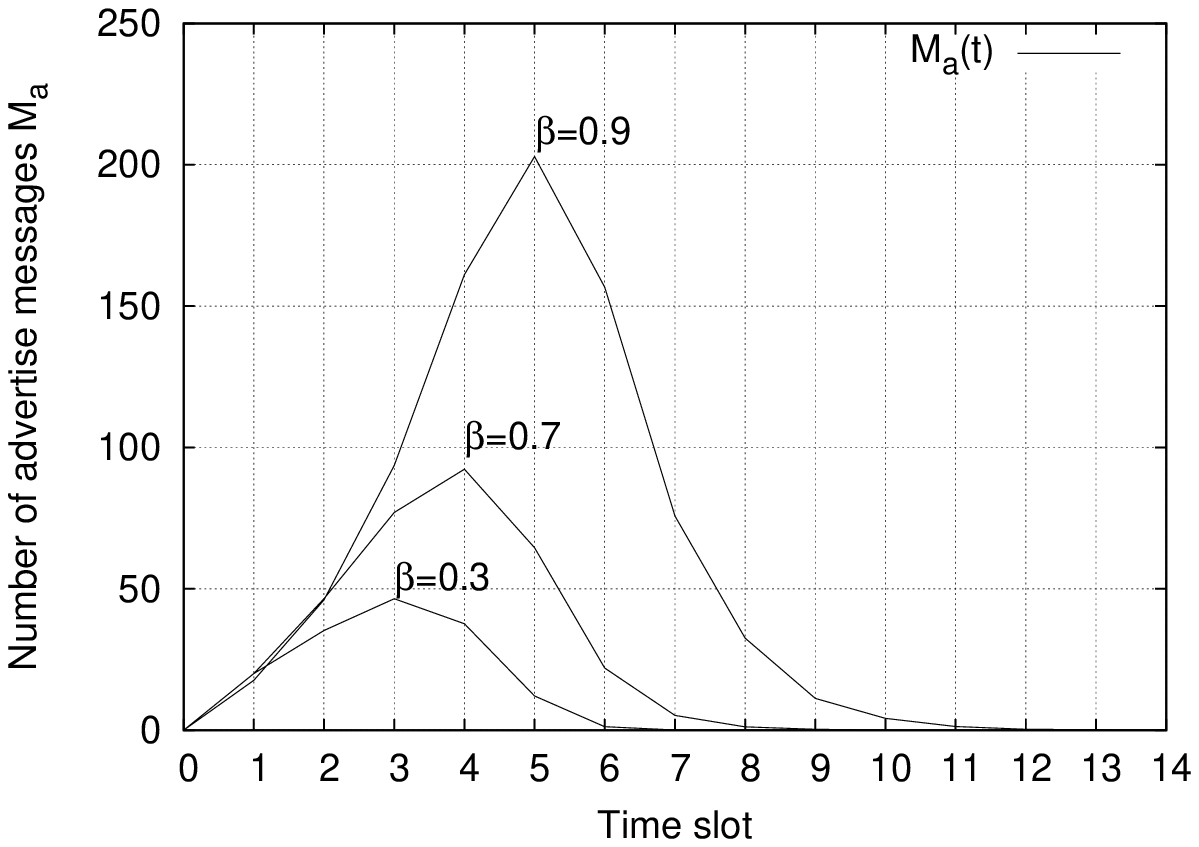}}
\caption{Overhead generated by adaptive probabilistic flooding.}
\end{center}
\end{figure*}

\subsection{Overhead}

Pure flooding generates a number of messages that is exponential in the number of nodes $N$. 
 Our adaptive probabilistic flooding algorithm is scalable in the sense that it generates  only $O(\N)$ messages, as shown below, at the expense of  $O(\N)$ counters, cf.~(\ref{eq:p}).

Consider a single advertisement from some source node $s$, and consider some relay node $j$ with degree $\delta$. The first time node $j$ receives a \FW\ message originated by $s$, it sends a copy on each output link, except the one where the \FW\ message has been received. This generates  $\delta-1$ messages. The second time $j$ receives a \FW\ message from the same source, it will forward the message over each of the $\delta-1$ links with probability $\Be<1$; so, at second reception, node  $j$ generates $(\delta-1)\beta$ messages on average. 
Iterating and taking into account $N$ advertisement processes (one per each node), we bound the total number of control messages \nm\ that are seen by the average node:
\begin{eqnarray}
\nm 	&\leq&	\N[(\degr-1)+(\degr-1)\Be+(\degr-1)\Be^2+\cdots] \nonumber \\
	&=&	\N(\degr-1)\sum_{n=0}^{\infty}\Be^n	\nonumber \\
	&=&	\N(\degr-1)\frac{1}{1-\Be}
\label{eqn:upperad}
\end{eqnarray}
(in other words, the whole network will carry $N\nm$ messages for a full round of $N$ advertisement processes). Note  that  (\ref{eqn:upperad}) is an upper bound since we do not account for  the detection of loops, which  reduces  the actual number of transmitted messages. While it is in principle  possible to refine the bound by considering the probability that loops form, this can  be done in closed form only for simple topologies such as random graphs~\cite{janson94graphs}.
Moreover, it turns out that this simple and conservative bound  matches very well, as we will see, the empirical results found by simulation.  

Note also that, as the first flood is always performed, convergence of the primary \PA\ path to the shortest path is always guaranteed. Hence, the backoff parameter \Be\ affects only the quality of the  secondary path \SE: by tuning \Be,  we can upper bound the algorithm overhead \nm\  while  matching the required level of path quality.

Now let us focus on the number of messages generated during the advertisement process.
Fig.~\ref{fig:msg} depicts, as a function of \Be, the upper bound (\ref{eqn:upperad})  along with the number of messages  \nm\ gathered by means of simulation, for the four topologies early outlined.   Note that the upper bound
associated with  Abilene network is computed using the fitted parameter values $(N,\degr)=(11,2.6)$, while for the other 22-nodes networks, only the upper bound for the Tiger2 topology is shown $(N,\degr)=(22,3.6)$ to avoid cluttering the picture. It can be noted that the overall number of messages generated by the advertisement procedure triggered by all nodes in the network remains low, topping to a few hundreds for high values of \Be. Also, notice that the number of messages generated in all the real networks is very similar, with Tiger2 as the worst case. This shows the robustness of our algorithm, given the various  topological properties of the considered networks.

Although the overall number of messages remains low for any value of \Be, we miss a reference for comparison: more insights can be gathered  from Fig.~\ref{fig:overhead}, that compares the number of messages generated by the advertise procedure with the periodic keep alive messages due to the path refresh function. Considering a fixed observation time window of $W=1$\,s, we vary the \tadvertise\ period in $[0.1,1]$\,s interval and set the refresh time to the typical value of \trefresh=50\,ms. We then compute the overall number of refresh messages \nrefresh\ by taking into account the actual primary  \PA\ and secondary \SE\ paths lenghts gathered in simulation  as $\nrefresh=\frac{W}{\trefresh}(length(\PA)+length(\SE))$. We finally gauge the number of advertisement messages in simulation for the Tiger2 network, that represents the worst case, so as to gather conservative results.
 As the picture clearly shows, the overhead due to the advertise procedure is very low even for very small durations of the advertise period \tm\ (well below 1\,sec), and even for large values of $\Be$: for $\tm\ge 1$\,sec the overhead becomes clearly negligible for any $\beta\le 0.9$.

Finally, Fig.~\ref{fig:msgt} depicts the evolution over (slotted) time of the number of messages carried over the Tiger2 network during an advertisement originated by a single source: we observe that, after some initial exponential growth due to the flooding process, the backoff factor kicks in and slows down the growth, which then dies out very fast, due to an increasing number of flooding paths being probabilistically cut out. This auto-termination feature is a very desirable property of the algorithm, and further suggests that advertisement periods do not need to overlap, but can rather be \emph{staggered}.
This could be achieved either with a simple policy (e.g., periodically at random within $[0,2\tadvertise]$) or with more sophisticated schemes (e.g., each node deciding autonomously whether to trigger a new advertisement depending on the measured control messages load).  In turn, this implies that instead of keeping $O(N)$ counters (one for each source in case of advertisements \emph{in parallel}), the system could perform advertisement \emph{in series} and keep a small number of $O(1)$ counters (using a modulo function to solve unlikely contentions due to independent simultaneous triggering of advertisement processes by multiple nodes). This is an interesting direction for future research, that we aim at pursuing in the following.

\begin{figure*}[t]
\begin{center}
\subfigure[\label{fig:connectivity}Connectivity probability of primary and secondary paths]{\includegraphics[width=0.3\textwidth]{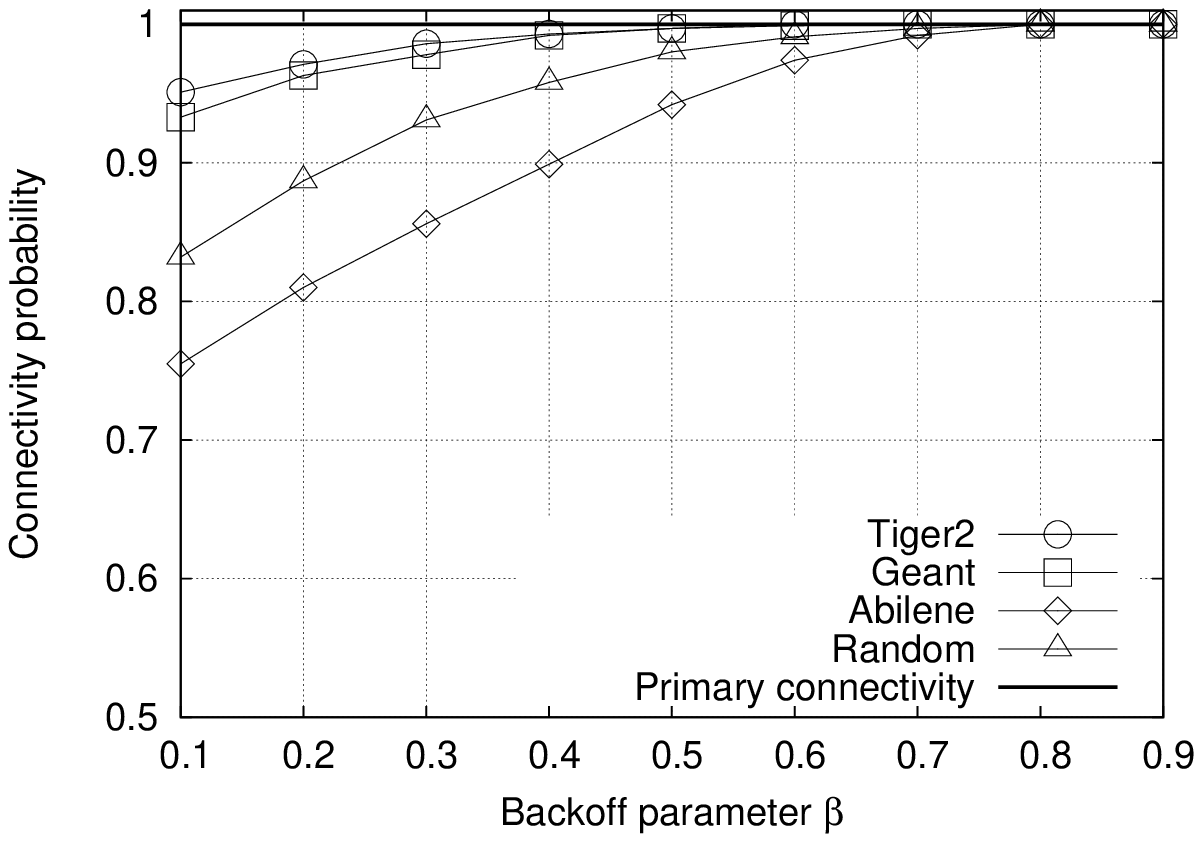}}
\subfigure[\label{fig:optimality}Optimality probability  of primary and secondary paths ]{\includegraphics[width=0.3\textwidth]{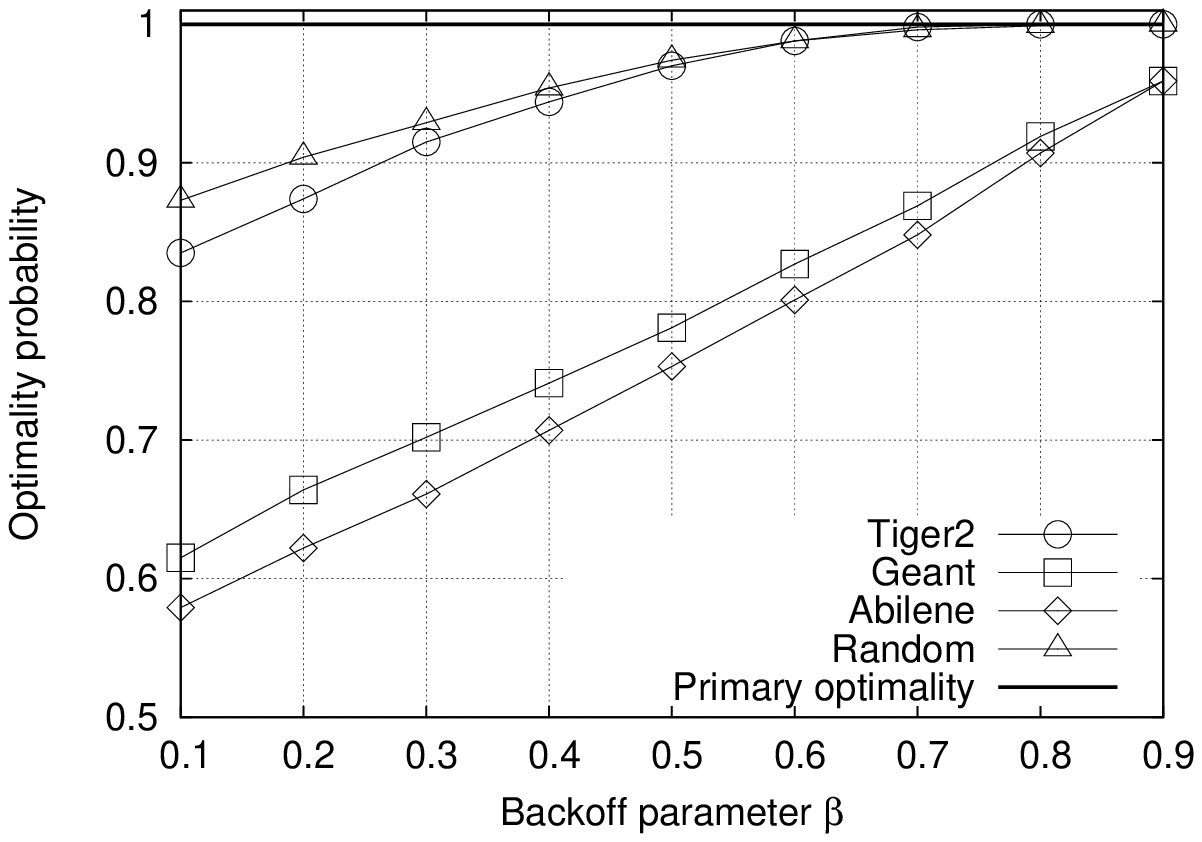}}
\subfigure[\label{fig:overlap}Average overlap between primary and secondary path $\| \PA \cap \SE\| $ for non-optimal seconday paths overlap]{\includegraphics[width=0.3\textwidth]{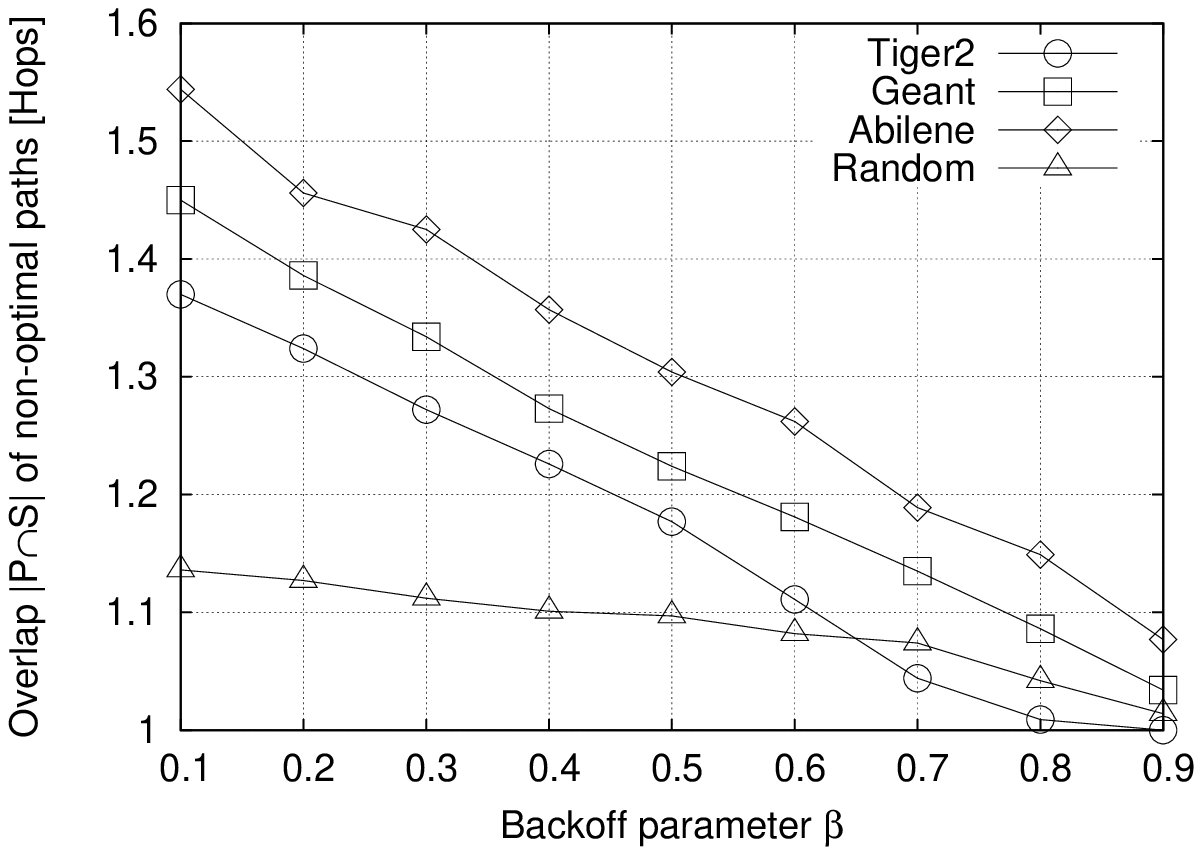}}
\caption{Quality of the paths found by adaptive probabilistic flooding.}
\end{center}
\end{figure*}

\subsection{Path quality}\label{sec:Experiments}
Let us now focus on the quality of the paths that the adaptive probabilistic flooding algorithm is able to find. For simplicity, we let each node advertise itself once at time $t=0$ and evaluate the connectivity and optimality of the primary and secondary paths. Results are averaged over 10 simulations over the real topologies, and over 50 graph instances in the synthetic Random graph case.

Fig.~\ref{fig:connectivity} depicts the \emph{connectivity} probability of the primary and secondary paths as a function of \Be: as expected, primary connectivity does not depend on \Be\ and is always guaranteed. Since  a primary path is always found, the connectivity index is relevant for the secondary path only:  we see that all secondary paths are connected in all networks when $\Be \ge 0.7$ (which correspond to small overhead in Fig.~\ref{fig:overhead}).

Fig.~\ref{fig:optimality} reports the \emph{optimality} probability
of the primary and secondary paths as a function of \Be: again, since  the shortest path is always eventually found, the optimality of the primary path is guaranteed. Thus, the optimality index is relevant only  for the secondary path:  we see that a significant percentage (from 60\% to 85\%, depending on the topology) of secondary paths are optimal even for a very low value of $\Be=0.1$, and that at least 90\% of secondary paths are optimal for all considered topologies when $\Be\ge 0.8$. Moreover, we observe  that optimality gracefully degrades \Be, and furthermore with similar (roughly linear) slope across all topologies.
This is a desirable behavior: as no phase transition nor knee appear in the path quality slopes, tuning \Be\ between low overhead (low \Be) vs high path quality (high \Be) is not critical.

Finally, we dissect the reason behind the sub-optimality of some secondary  paths. Recall that a secondary path is optimal if it is the shortest and most diverse path compared to the primary. Hence, sub-optimality of the secondary path may be due to  either (i) a non-zero \emph{overlap} between primary and secondary paths,  $| \PA \cap \SE |>0$,  or (ii) a path with a stretch  over the optimal secondary path  larger than one $\lenS/L_{s'} > 1$.  Fig.~\ref{fig:overlap}  depicts the overlap, i.e., the number of nodes that primary and secondary paths have in common, conditioning over the sub-optimal paths (i.e., the overlap of optimal secondary paths is not accounted for in the picture). As shown by the figure,  sub-optimality seems to be tied to slightly more than one node in common as  $|\PA \cap \SE|\in [1,1.5]$. Furthermore, as the average overlap is always $| \PA \cap \SE | \ge 1$ for any \Be, we can conclude that overlapping paths are significantly more common that long-stretching paths.

\section{Conclusions}\label{sec:Future}
We have presented a novel flooding based algorithm for multiple-path discovery: the algorithm trades a small amount of state in routers, i.e., $O(N)$ counters, in order to significantly limit the number of messages generated by flooding through an adaptive probabilistic algorithm. 

Simple analytical bounds, confirmed by simulation results, show the overhead entailed by the advertisement procedure to be low (with respect to the amount of keep-alive messages needed to keep the path up-to-date) and auto-terminating (due to the multiplicative decrease of the flooding probability).

Simulation results also testify excellent performance in terms of path quality:
connectivity and optimality of the primary path are achieved by design, while 90\% of secondary paths are also optimal when $\Be\ge 0.8$ (or otherwise decrease linearly for lower \Be).  Interestingly, the low percentage of low path is due to a very limited amount of share of faith between paths (1.5 nodes in the worst case, for $\Be=0.1$).

As part of our future work, we want to carry on more realistic experiments on a wider set of topologies and further reduce the amount of state to $O(1)$.

\section{Acknowledgement*}
This work was funded by Celtic TIGER2 project.

\bibliographystyle{IEEEbib}
\bibliography{biblio}
\end{document}